# Antiferromagnetic opto-spintronics: Part of a collection of reviews on antiferromagnetic spintronics

P. Němec,[1] M. Fiebig,[2] T. Kampfrath,[3] and A. V. Kimel[4]

*[1]Faculty of Mathematics and Physics, Charles University, Ke Karlovu 3, 121 16 Prague 2, Czech Republic*
*[2]Department of Materials, ETH Zurich, Vladimir-Prelog-Weg 4, 8093 Zurich, Switzerland*
*[3]Department of Physical Chemistry, Fritz Haber Institute of the Max Planck Society, Faradayweg 4-6, 14195 Berlin, Germany*
*[4]Radboud University, Institute for Molecules and Materials, Heyendaalseweg 135, 6525 AJ Nijmegen, The Netherlands*

**Control and detection of spin order in ferromagnets is the main principle allowing storing and reading of magnetic information in nowadays technology. The large class of antiferromagnets, on the other hand, is less utilized, despite its very appealing features for spintronics applications. For instance, the absence of net magnetization and stray fields eliminates crosstalk between neighbouring devices and the absence of a primary macroscopic magnetization makes spin manipulation in antiferromagnets inherently faster than in ferromagnets. However, control of spins in antiferromagnets requires exceedingly high magnetic fields, and antiferromagnetic order cannot be detected with conventional magnetometry. Here we provide an overview and illustrative examples of how electromagnetic radiation can be used for probing and modification of the magnetic order in antiferromagnets. Spin pumping from antiferromagnets, propagation of terahertz spin excitations, and tracing the reversal of the antiferromagnetic and ferroelectric order parameter in multiferroics are anticipated to be among the main topics defining the future of this field.**

I. INTRODUCTION

Antiferromagnets (AFs) are materials in which the elementary magnetic moments are spontaneously long-range ordered, but with a net magnetic moment that is zero or small compared to the sum of the magnitudes of the participating magnetic moments[1]. AFs are considerably more common than ferromagnets (FMs). In particular, they are permitted in each magnetic symmetry group, in contrast to ferromagnetism. They can be insulators, metals, semimetals, semiconductors or superconductors, whereas FMs are primarily metals. In contrast to FMs, there is a large number of ways how magnetic moments can be arranged on a lattice to achieve a zero net moment[2,3]. For example, in perovskite $LaFeO_3$ all nearest-neighbour magnetic atoms are collinear and antiferromagnetically aligned whereas in $LaMnO_3$ alternatively aligned ferromagnetic planes are present[3]. In $YMnO_3$, there are three magnetic sublattices rotated by 120° (Ref. 4). In rare-earth metals[3] and multiferroics like $TbMnO_3$ (Ref. 5), helical order is present.

I.a Functional properties of antiferromagnets

AFs exhibit a variety of unique functionalities of either intrinsic or engineered nature. On the intrinsic side, antiferromagnetism may be linked to a small secondary magnetization or polarization. The magnetization can result from an antisymmetric form of exchange coupling termed Dzyaloshinskii-Moriya interaction[6,7]. It results in a canting of two initially collinear antiferromagnetic spin-sublattices by about 1°, corresponding to a magnetization of about ~$10^{-3}$ $\mu_B$, as discovered in the widespread mineral hematite ($\alpha$-$Fe_2O_3$). Materials exhibiting this effect are sometimes called "canted AFs" or "weak FMs"[8].

If the antiferromagnetic order breaks inversion symmetry, spin-spin interactions may give rise to a symmetry-compatible electric polarization of about 1-100 nC cm$^{-2}$. The combination of antiferromagnetic and ferroelectric order in the same phase makes the material a so-called multiferroic[9,10]. The secondary polarization can be employed to probe and control the antiferromagnetic order which is typically of the helical type[5].

On the engineering side, the so-called "synthetic antiferromagnets" (SAFs) form a class of materials with antiferromagnetically coupled layers of FMs. Fundamental studies of exchange interaction in SAFs enabled the discovery of giant magnetoresistance (GMR) in Fe-Cr-Fe trilayers[11,12] and a revolution in magnetic hard-disk industry for which the Nobel Prize in Physics was awarded in 2007 (for more deteiles see the article on SAFs in this focused issue). AFs are also frequently used as reference layer in spin valves which fixes the magnetization orientation in FM due to the exchange bias[13,14] appearing as unidirectional magnetic anisotropy[15].

However, the potential of AFs is significantly larger, as envisioned in the recent concept of antiferromagnetic spintronics[16-21] where antiferromagnetic materials serve as active parts of spintronic devices. In particular, antiparallel spin sublattices in AFs, which produce zero dipolar fields, imply the insensitivity to magnetic-field perturbations[22,23] and multi-level stability[24,25] in magnetic memories. Another appealing functional property of AFs is the orders of magnitude faster spin dynamics than in FMs [26]. The frequency of uniform spin precession (antiferromagnetic resonance) is in the terahertz (THz) range due to strong exchange interaction between the spin sublattices[27-29], whereas ferromagnetic resonance, which is given by the weaker anisotropy field, is in the gigahertz (GHz) frequency range[26,29]. However, the absence of a net magnetic moment, the small size of magnetic domains and the ultrafast magnetization dynamics make probing of antiferromagnetic order by common magnetometers or magnetic resonance techniques notoriously difficult.

I.b Optics on antiferromagnets

Light or, more general, electromagnetic radiation is an invaluable tool for probing and manipulating of magnetic order. Thus, established tools for optical access to ferromagnetic order have been developed and refined for over 150 years[30]. Because of the lack of macroscopic magnetization, antiferromagnetism is much harder to access by optical techniques, however. The respective approaches are quite young and still under development. Section II of this review is devoted to a survey of techniques for probing antiferromagnetism in the optical range and beyond.

Manipulation of magnetic order by electromagnetic radiation is the second challenge. The strength of an external magnetic field required to reorient spins in AFs scales with the exchange interaction and thus can reach tens or hundreds of Tesla. Remarkably, light was shown to be an efficient tool to control and detect spins and their ultrafast dynamics in magnetic materials[26,31,32]. Some of the mechanisms for the optical control, such as inertia-driven spin switching[33], least dissipative impulsive excitation of spins[34], excitation of THz nanomagnons at the edge of the Brillouin zone[35], and reducing the spin noise below the standard quantum limit[36] have so far been demonstrated only in AFs or are even possible exclusively in these materials. Chapters III and IV of this review will give an overview on manipulation of antiferromagnetic order by electromagnetic radiation, including ultrafast processes aiming directly at modification of the quantum-mechanical coupling between spins and macrospin precession phenomena.

II. DETECTION TECHNIQUES

Several detection techniques exist for probing antiferromagnetic order in bulk crystals, among which neutron diffraction plays the major role. However, in the case of epitaxial thin films of nanometer thickness, which form the building blocks of spintronic devices, and for ultrafast transient processes as well as for spatially-resolved detection, this technique is usually not applicable. Below we describe three distinct groups of detection techniques of antiferromagnetic order based on the interaction of electromagnetic radiation with AFs.

II.a Linear optical studies

Starting from the discoveries of Michael Faraday (1846), John Kerr (1877) and Woldemar Voigt (1899), magneto-optics (MO) has been established as an efficient probe of magnetic order[30,37,38] with high spatial[39] and temporal[26] resolutions. Assume that an antiferromagnet has just two equivalent magnetic sublattices represented by spins $\mathbf{S}_1$ and $\mathbf{S}_2$. It is convenient to introduce two orthogonal vectors of magnetization $\mathbf{M}=\mathbf{S}_1+\mathbf{S}_2$ and antiferromagnetism $\mathbf{L}=\mathbf{S}_1-\mathbf{S}_2$. Linear optical properties are described by the dielectric permittivity tensor $\varepsilon_{ij}$, which can be written as a sum of the antisymmetric and symmetric parts $\varepsilon_{ij} = \varepsilon^{(a)}_{ij} + \varepsilon^{(s)}_{ij}$, where $\varepsilon^{(a)}_{ij} = -\varepsilon^{(a)}_{ji}$ and $\varepsilon^{(s)}_{ij} = \varepsilon^{(s)}_{ji}$. According to the Onsager principle, $\varepsilon^{(a)}_{ij}$ can only be an odd function with respect to $\mathbf{L}$, $\mathbf{M}$ and their combinations, while the dependence of $\varepsilon^{(s)}_{ij}$ on the order parameters is even[40].

Many time-resolved MO studies have been performed in weakly absorbing canted antiferromagnets[33,41-48], where the probe of antiferromagnetism relies on a linear dependence of $\varepsilon^{(a)}_{ij}$ on the secondary magnetization $\mathbf{M}$. Large MO Kerr effect was predicted also for noncollinear AFs Mn$_3$X (X = Rh, Ir, Pt)[49]. Let us consider a Cartesian coordinate system with the $x$- and $y$-axes in the sample plane and the $z$-axis along the normal to the sample. If in the otherwise isotropic medium with relative dielectric permittivity $\varepsilon$ an external stimulus induces $\varepsilon^{(a)}_{xy}$, it will break the degeneracy between right-handed and left-handed circularly polarized light waves propagating in the direction of the $z$-axis. Waves of opposite helicity will be refracted and absorbed differently resulting in circular birefringence and dichroism, respectively[50]. Due to the circular birefringence, linearly polarized light propagating along the $z$-axis, where $\mathbf{M}$ is oriented, will experience a polarization rotation by an angle

$$\alpha_F = \frac{\omega d}{2c\sqrt{\varepsilon}} \varepsilon_{xy}^a, \tag{1}$$

where $\omega$ is the frequency of the light wave, $d$ is the propagation length in the sample, and $c$ is the speed of light in vacuum. Magnetization-induced polarization rotation (or elipticity change) upon propagation or reflection are called the magneto-optical Faraday and Kerr effect, respectively. Even if the equilibrium magnetization is zero and no magnetic field is applied, the magnetization in AFs can still be induced by coherent dynamics of spins $\mathbf{M} \sim \mathbf{L} \times d\mathbf{L}/dt$ (Ref. 51). Therefore, the dynamic magnetization can also be used to probe pump-induced changes of the spin ordering in AFs[4,27,28,52].

An alternative approach for MO detection is to use $\varepsilon^{(s)}_{ij}$, which depends quadratically on $\mathbf{L}$ and is, therefore, present even in compensated AFs (where M = 0). If in an otherwise isotropic medium the spins get spontaneously and antiferromagnetically ordered along the *x*-axis, the alignment will change $\varepsilon^{(s)}_{ij}$, so that $\varepsilon^{(s)}_{xx} \neq \varepsilon^{(s)}_{yy}$, inducing linear dichroism and birefringence[26,30,37,39,53-56]. The former, which is usually called the Voigt effect, can be measured as a polarization rotation[57]. Contrary to the Faraday effect, this polarization rotation angle $\alpha_V$ is the largest when the light propagation direction is perpendicular to the spin orientation. It also depends on the angle between the polarization of the incoming light ($\beta$) and $\mathbf{L}$ orientation ($\varphi$) according to

$$\alpha_V(\beta) = P^{MLD} \sin 2(\varphi - \beta). \tag{2}$$

The MO coefficient $P^{MLD}$ scales quadratically with the $\mathbf{L}$ projection onto the plane perpendicular to the light propagation direction. It is connected with the magnetic linear dichroism (MLD): $P^{MLD} = 0.5(T_\parallel / T_\perp - 1)$, where $T_\parallel$ and $T_\perp$ are the amplitude transmission coefficients for light polarized parallel and perpendicular to $\mathbf{L}$, respectively[54,57]. Experimentally, the quadratic dependence on the order parameter and the low sensitivity of AFs to external magnetic fields significantly complicate the separation of the magnetic-order-related signal from linear dichroism and birefringence of other origins (e.g., strain- or crystal structure-related)[58]. This problem can be circumvented if a two-beam pump-probe detection scheme is used that utilizes the light-induced modulation of antiferromagnetic order [see Figs. 2(a) – (c)][57]. For AFs, the quadratic MO effects were used for the detection of pump-induced dynamics both in weakly absorbing insulators[4,34,35,45,46,52] and in metals[57].

The quadratic dependence of $\varepsilon^{(s)}_{ij}$ on $\mathbf{L}$ results also in the polarization-independent phenomena of magneto-refraction and magneto-absorption. The effects were observed as an intensity change of the diffracted X-rays in $La_{0.5}Sr_{1.5}MnO_4$ (Ref. 59), transmitted or reflected light in EuTe (60), FeRh (Ref. 61) and $Cr_2O_3$ (Ref. 62). Finally, also stimulated Raman scattering on magnons, employed for probing the strength of the exchange interaction in $KNiF_3$, can be explained in terms of a quadratic coupling to $\mathbf{L}$, assuming that spins oscillate at magnon frequencies[63].

MO effects linear with respect to **L** can be observed in AFs for which both time-reversal and space-inversion symmetries are broken, but the combined symmetry operation is retained. In this case, spatial dispersion implies an additional contribution to $\varepsilon^{(s)}_{ij}$ proportional to **k**·**L**, where **k** is the wave vector of light[64]. Alternatively, light waves travelling in opposite direction may be transmitted with different intensity and polarization, an effect termed nonreciprocal directional dispersion. As an extreme case, unidirectional propagation of visible light has been demonstrated in $CuB_2O_4$ (Ref. 65).

MO effects of similar phenomenology are present also in X-ray[53,66-70] and THz[71,72] spectral ranges. For instance, the effects quadratic with respect to **L** are rather strong in the soft X-ray range due to the resonant enhancement occurring at the 2p edges of 3d-transition metals and the 3d edges of rare-earth elements[53]. Therefore, they allow for element-selective investigations of magnetic properties of constituents of magnetic materials as well as for the element-selective imaging of magnetic domain structures[53,70]. Moreover, a combination of X-ray magnetic linear dichroism (XMLD) with photoemission electron microscopy (PEEM) enables direct imaging of antiferromagnetic domains[73-77] with a spatial resolution below 100 nm. This technique was used, for example, for visualisation of current-induced changes in the domain structure in the CuMnAs memory device [see Fig. 1(d)-(j)][77]. In certain cases, element specificity and/or in-depth resolution can be achieved also in the optical wavelength range[78-81].

II.b Nonlinear optical studies

For the MO effects described in the previous section, the polarization rotation experienced by the probing light beam is independent of the amplitude of the light field. Therefore, these phenomena belong to the realm of linear optics. However, at sufficiently high light intensities, as provided by lasers, nonlinear optical effects begin to manifest[82] which significantly broadens the portfolio of material properties that can be probed. Among nonlinear effects quadratic in the electric field **E** of the incident light wave, second-harmonic generation (SHG) plays a very important role. This effect, where two photons from the incident wave at frequency $\omega$ are anihilated and one photon at frequency $2\omega$ is created, is typically described by the equation **P**$(2\omega) \propto$ **E**$(\omega)$**E**$(\omega)$. This electric-dipole-only process is restricted to noncentrosymmetric materials[82]. However, magnetic-dipole or electric-quadrupole contributions like **M**$(2\omega) \propto$ **E**$(\omega)$**E**$(\omega)$ or **P**$(2\omega) \propto$ **E**$(\omega)$**H**$(\omega)$ may have to be included[83,84].

Being the lowest-order nonlinear optical process, the efficiency of SHG is rather high. It can be used for distinguishing materials or ferroic states of different symmetry, including probing of the magnetic order of materials, as illustrated in Figs. 1(a)-(c)[85]. This possibility started to be studied theoretically in the 1960s[83,86,87] and experimentally in the 1990s[84,88-92]. Nowadays, SHG is a very powerful method that enables to discriminate magnetic space groups, which are indistinguishable to diffraction methods, or to reveal hidden magnetic-field-induced phase transitions[85]. Magnetic SHG is especially powerful in the case of AFs where linear MO effects are not present, as discussed above. By polarization-dependent SHG spectroscopy, the antiferromagnetic contribution to the SHG intensity can be identified and measured background-free as depicted in Fig. 1(b). Using the interference of the signal field with a reference field, it can also be used for a visualization of 180° domains in AFs[85,93-96] as shown for $Cr_2O_3$ in Fig. 1(c). Alternatively, the magnetic SHG can be employed as a sensitive tool to probe the pump-induced dynamics in AFs[31,96-100].

Since each form of order breaks symmetry in a different way[101], magnetic and ferroelectric order can be readily separated via the different polarization of the corresponding SHG contribution emerging at the respective ordering temperature[85]. SHG is, therefore, ideal for probing the coexistence and coupling of ordered states and domains in multiferroics in a single experiment[102].

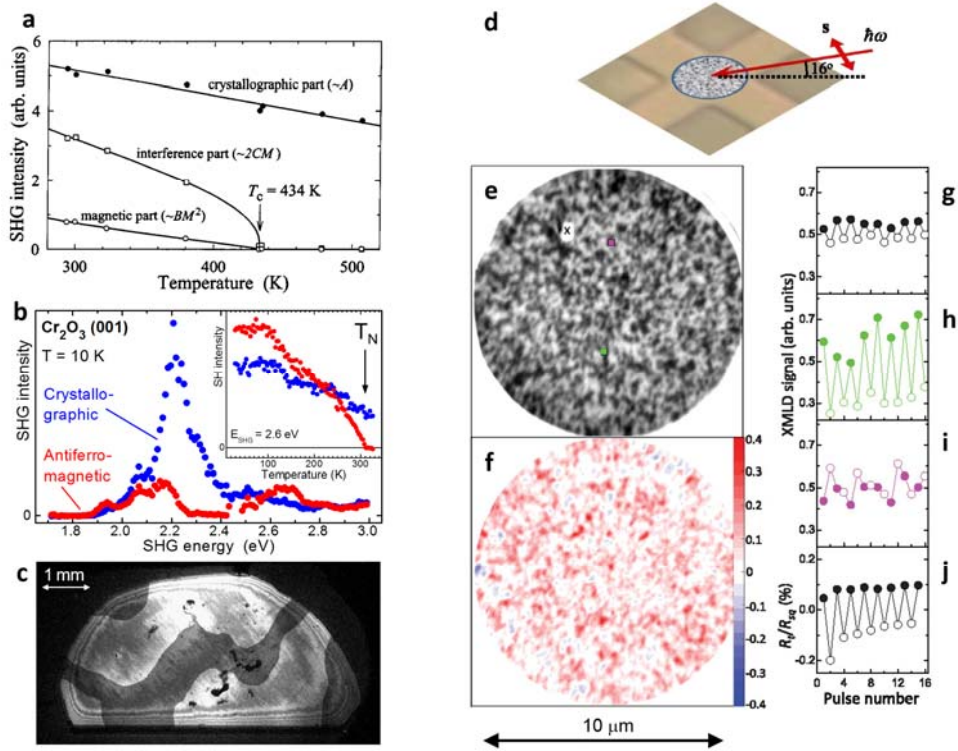

**Figure 1 Visualization of antiferromagnetic domains by SHG (a-c) and XMLD-PEEM (d-j).**
**a**, Temperature variation of crystallographic, magnetization-induced, and interference terms of the SHG intensity of the Bi-YIG/GGG(111) film. **b**, SHG spectrum of a *z*-oriented $Cr_2O_3$ platelet. Light polarizations are chosen such that the crystallographic and antiferromagnetic SHG contribution are separated and measured background-free. The inset shows that the antiferromagnetic contribution vanishes at the Néel temperature whereas the crystallographic contribution remains[84]. **c**, SHG image of antiferromagnetic 180° domains on the $Cr_2O_3$ sample. By interference of the crystallographic and antiferromagnetic SHG contributions, opposite domain states exhibit a different degree of brightness[85,94].
**d**, Geometry of XMLD-PEEM measurements of a device made from CuMnAs. X-rays are incident at 16° to the surface, with the polarization vector **s** in the film plane. **e**, XMLD-PEEM image of the central section of the device. **f**, Difference between XMLD-PEEM images taken after applying trains of alternate orthogonal current pulses of 6.1 MAcm$^{-2}$. **g**, Spatially averaged XMLD signal after each pulse train. Open and filled symbols represent the two orthogonal pulse directions. **h**, **i**, As for **g**, but for the 200 × 200 nm$^2$ regions marked by green and pink squares in **e**, respectively. **j**, Change in the transverse resistance following the same pulse sequence. Figure reproduced with permission from: **a**, Ref. 92; **d-j**, Ref. 77.

II.c THz emission and transmission spectroscopy

Recent technological advances in the generation of THz radiation motivated a growing interest in this spectral region[103-111]. In contrast to optical radiation, which predominantly interacts with valence electrons, THz radiation couples to various low-energy excitations such as molecular rotations, lattice vibrations and spin waves. If magnetization precession is induced, for example by the impact of an ultrashort laser pulse[26,31,32], the oscillating magnetic dipoles lead to the emission of an electromagnetic wave. For a thin layer with uniformly oscillating magnetization $\mathbf{M}(t)$ on top of a thick substrate with refractive index $n_s$, the resulting electric field $\mathbf{E}(t)$ in the air half-space directly behind the magnetic layer can be expressed as[112]

$$\mathbf{E}(t) = \frac{Z_0 n_s d}{(n_s + 1)c} \mathbf{e} \times \frac{\partial \mathbf{M}}{\partial t}. \quad (3)$$

Here, $Z_0 \approx 377\ \Omega$ is the vacuum impedance (which is assumed to be much smaller than the resistance of the magnetic film), and $\mathbf{e}$ is the unit vector normal to the layer, whose thickness $d$ is assumed to be much smaller than the THz wavelength within the layer. As the resonant frequencies of uniform spin precession in AFs are typically as high as several THz[26,31,32], the emitted radiation is in the THz spectral range also. Consequently, it is possible to study the projection of the magnetization trajectory on the plane perpendicular to $\mathbf{k}$ by measuring both linear polarization components of the THz wave that is emitted by the optically excited material[113-116]. Therefore, THz emission spectroscopy enables to study ultrafast magnetization dynamics *directly* (and sometimes even with element specificity[116]). This feature is in contrast to MO[117,118] and SHG[119] experiments where the pump-induced changes in the corresponding coefficients may influence the measured signal dynamics, especially on sub-picosecond time scales. Note that illumination with a resonant THz pulse can also induce the spin precession which can, in turn, be detected by measuring the reemitted THz radiation. This configuration is tantamount to a THz transmission experiment and has been routinely applied to characterize long-wavelength antiferromagnetic magnons[120,121].

III. ULTRAFAST MODIFICATION OF MAGNETIC ORDER

The research on ultrafast dynamics of magnetic order is a challenging area in the physics of magnetism. The development of femtosecond lasers has opened the way to create external stimuli that are shorter than fundamental time scales such as spin-lattice relaxation or precession time, thereby allowing for an extremely fast manipulation of the magnetic state of materials[26]. As detection of antiferromagnetic order is difficult, these phenomena were originally studied mainly in FMs[26]. Nevertheless, also in AFs there are reports about ultrafast changes of the antiferromagnetic order which include its quenching to a paramagnetic state, switching to a ferromagnetic state and a spin reorientation.

The first theoretical prediction of ultrafast quenching (demagnetization) of antiferromagnetic order was reported for NiO in Ref. 122. Experimentally, the optically induced phase transition from the antiferromagnetic to the paramagnetic state was observed in the canted antiferromagnet iron borate $FeBO_3$ in Ref. 41. Unlike in metallic nickel[123], where the photoexcitation directly heats up the electronic system, the magnetic order quenching in dielectric $FeBO_3$ is caused by an increase of the magnon temperature due to energy transfer from the heated lattice[41]. The quenching of antiferromagnetic order in dielectric $Cr_2O_3$ was reported in Ref. 99 and in dielectric half-doped manganite $La_{0.5}Sr_{1.5}MnO_4$ in Ref. 59. This phenomenon can be also used for a detailed magnetic characterization of AFs, as demonstrated for a thin film of metal CuMnAs [see Fig. 2(a)-(c)][57].

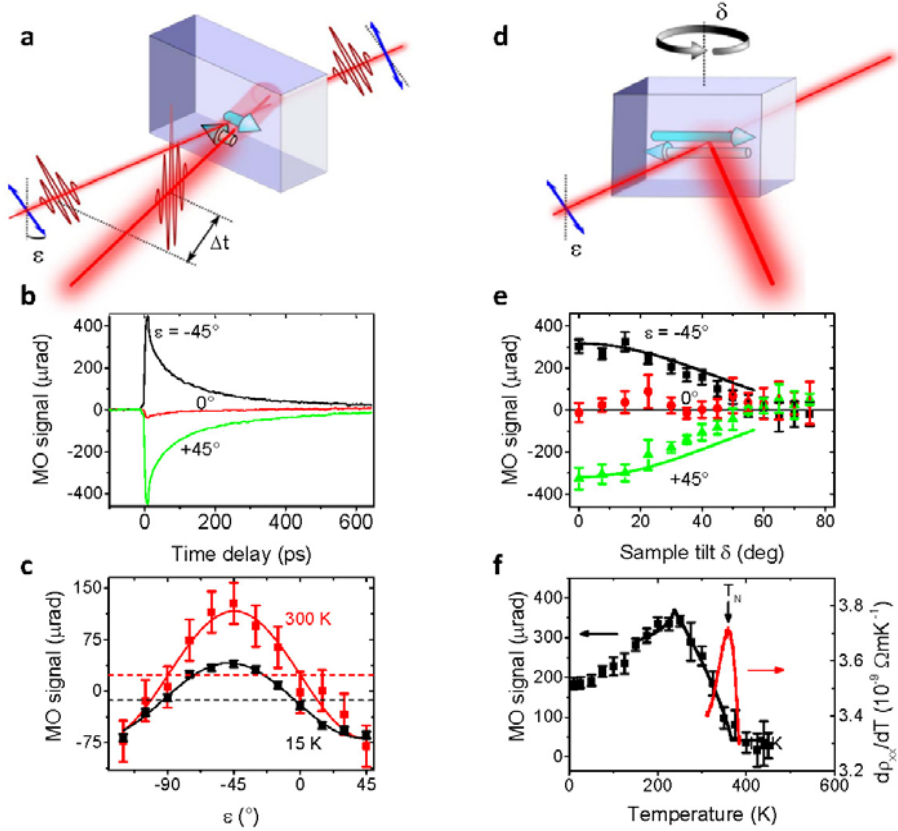

**Figure 2 Determination of uniaxial magnetic anisotropy direction and Néel temperature from pump-induced demagnetization in CuMnAs film**. **a**, Schematic illustration of the pump-induced reduction of a light polarization plane rotation due to the Voigt effect. **b**, Pump-induced change of MO signal measured for various probe polarization orientations, $\varepsilon$, as a function of time delay, $\Delta t$, between pump and probe pulses. **c**, Probe-polarization dependence of MO signal measured for $\Delta t = 60$ ps at 15 K (black points) and 300 K (red points) for a 10 nm thick CuMnAs film. Solid lines are fits by Eq. (2) plus polarization-independent backgrounds (dashed horizontal lines). **d**, Schematic illustration of sample tilting around an axis perpendicular to the direction of the magnetic moments leading to a reduction of moments projection onto plane perpendicular to the probe light propagation direction. **e**, MO signal measured for $\Delta t = 60$ ps (points) as a function of the sample tilt around the [1-10] substrate direction at 300 K. Solid lines depict the function $\cos^2 \delta$ describing the moment projection reduction expected for the situation shown in **d**. **f**, Temperature dependence of MO signal (points) and sample resistivity temperature derivative (red line). The vertical arrow indicates the Néel temperature $T_N$. Figure reproduced with permission from Ref. 57.

The dynamic change of antiferromagnetic to ferromagnetic order due to ultrashort external excitations was studied in detail in metallic FeRh, which undergoes a first-order magneto-structural transition from an antiferromagnetic to ferromagnetic phase around 380 K [see Fig. 3(a)]. As reported independently in Refs. 124 and 125, the illumination of FeRh films by femtosecond optical pulses drives the antiferromagnetic-to-ferromagnetic transformation on a picosecond time scale [see Fig. 3(f)][126]. Subsequent studies revealed[126-130] that the observed behaviour is due to an interplay between the local magnetic moment, nonlocal domain growth and/or alignment, and magnetization precession launched by the varying demagnetizing fields during the build-up of ferromagnetic order in the illuminated film [see Figs. 3(b)-(e)][126].

Finally, laser-induced ultrafast reorientation of spins in AF was reported in the dielectric orthoferrite $TmFeO_3$ where a temperature-dependent magnetic anisotropy is observed at 80–91 K [see Figs. 3(j)-(k)][42]. Similar temperature-related effects were reported for $DyFeO_3$ (Ref. 131), where around 39 K a reorientational magnetic phase transition takes place, and in Ref. 132 for CuO (Ref. 132), where around 220 K a transition from a collinear antiferromagnetic structure to a multiferroic phase with a spiral spin structure occurs. In NiO, the ultrafast reorientation of $Ni^{2+}$ spins due to the shift of $3d$ orbital wave functions, which accompanies the pump-excitation of $d$-$d$ transitions, was reported in Ref. 97. Similarly, the transient change in the easy direction was proposed to lead to a rapid photo-mediated reorientation of $Ni^{2+}$ spins from the local $[11\bar{2}]$ direction to the local $[111]$ direction in NiO (Ref. 31). Another example of such ultrafast modification of magnetic order is the transition from an insulating antiferromagnetic phase to a conducting metallic phase that happens in the perovskite manganite $Pr_{0.7}Ca_{0.3}MnO_3$ (Refs. 31 and 133). The pump-induced changes of magnetic interactions were studied in detail in bilayers AF/FM (NiO/NiFe in Ref. 134 and 135, $FeF_2$/Ni in Ref. 136, IrMn/Co in Ref. 137, CoO/Fe in Ref. 138, and CuMnAs/Fe in Ref. 139). It was concluded that the dynamical pump-probe experiments can provide markedly distinct information about the bilayer properties than the equilibrium measurements of magnetic hysteresis loops.

The laser-induced ultrafast reorientation of spins in AFs described above is closely connected with spin switching when an external stimulus (e.g. strong laser pulse) induces a transition between metastable magnetic states. The fastest way to reorient magnetization **M** in FMs without demagnetization is through precessional motion in an applied field, which can be described by the Landau-Lifshitz-Gilbert equation. This equation is of first order with respect to time and, therefore, does not contain inertial terms such as acceleration[33]. In other words, the field has to be applied until the magnetization crosses the potential barrier that is separating one minimum from another. In contrast, the dynamics of AFs is described in terms of the motion of the antiferromagnetic unit vector **L** whose equation of motion is of second order with respect to time and, thus, shows inertia-like motion[33] [see Fig. 4(a)]. Consequently, in AFs it is possible to achieve spin switching even if the external stimulus is shorter than the time needed for overcoming the barrier as demonstrated in $HoFeO_3$ [see Figs. 4(b)-(c)][33]. This inertial character of coordinated dynamics of two interacting sublattices is also at the core of the AFs switching mechanism by THz pulses which was theoretically proposed in Ref. 140.

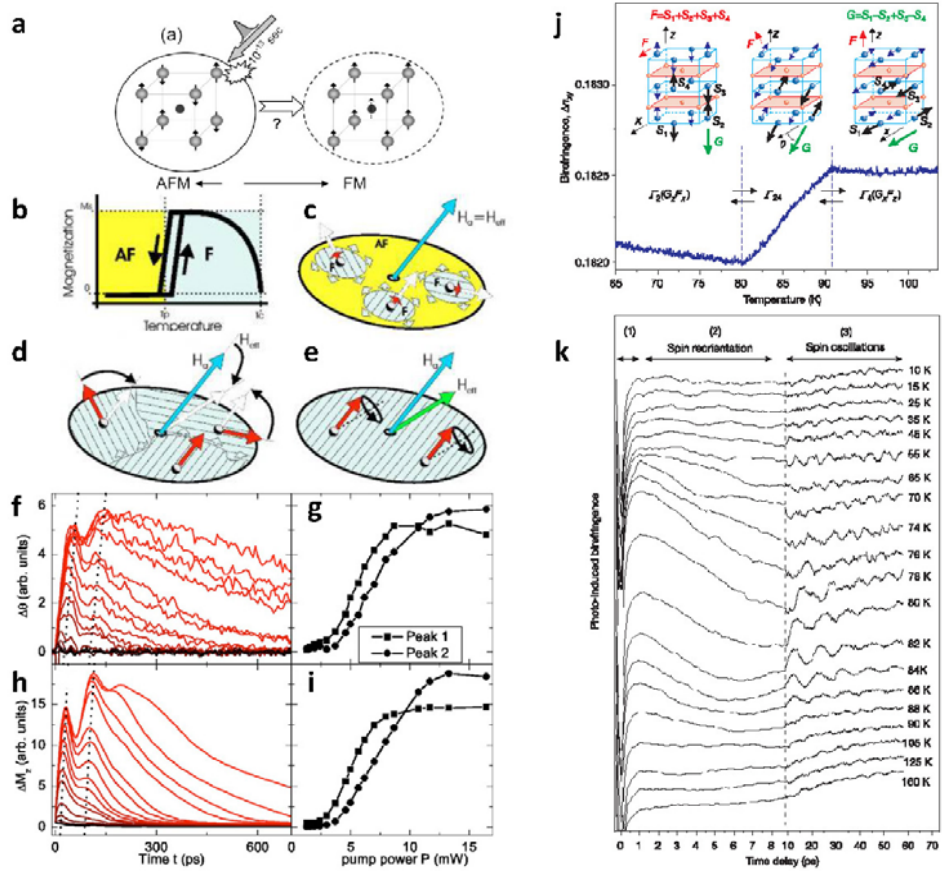

**Figure 3 Ultrafast modification of magnetic order in FeRh (a-i) and TmFeO$_3$ (j-k).**
**a**, Schematic of the ultrafast generation of ferromagnetic order by inducing an antiferromagnetic -to-ferromagnetic transition in FeRh when excited with femtosecond optical pulses. **b**, At low temperatures, FeRh is antiferromagnetic (yellow) with local iron moments $m_{Fe} = 3\mu_B$ and no appreciable moment on rhodium. At elevated temperatures, the system is ferromagnetic (green) with local iron and rhodium moments of $m_{Fe} \approx 3\mu_B$ and $m_{Fe} \approx 1\mu_B$. **c**, The growth of local magnetization and areal growth of the ferromagnetic phase is followed by a growth of net magnetization by alignment of individual domains. **d**, The demagnetization field grows equally, leading to a canting total effective field. **e,** The homogeneous magnetization starts precessing around the new effective field. Red arrows represent the local magnetization **M**(r,t). **f**, Transient Kerr rotation measured at 3.5 kG as a function of time; applied pump power varies from 1.3 mW (lowest curve) to 16.5 mW (highest curve). **g**, Height of the first and second peaks plotted against the incident pump power. **h**, **i**, Corresponding simulations[126].
**j**, Linear optical birefringence in TmFeO$_3$ as a function of temperature. Insets show the corresponding arrangement of spins: below 80 K the spins lie nearly along the ±z axes while above 91 K the spins are slightly canted from the ±x axes. **k**, Excitation and relaxation of the antiferromagnetic moment measured via changes in the magnetic birefringence. When the pump laser pulse is absorbed via the excitation of the localized electronic states of the Fe$^{3+}$ and Tm$^{3+}$ ions, the following relaxation process can be observed. First, the excitation decays via phonon cascades and the phonon system thermalizes with 0.3-ps relaxation time [process (1)]. The phonon–phonon interaction sets a new lattice temperature so that the equilibrium anisotropy axis is changed. Consequently, the resulting motion of spins to the new equilibrium happens with 5-ps response time [process (2)] which is followed by oscillations of moments around their new equilibrium with an approximate 10-ps period. [process (3)]. Figure reproduced with permission from: **a**, Ref. 124; **b-i**, Ref. 126; **j-k**, Ref. 42.

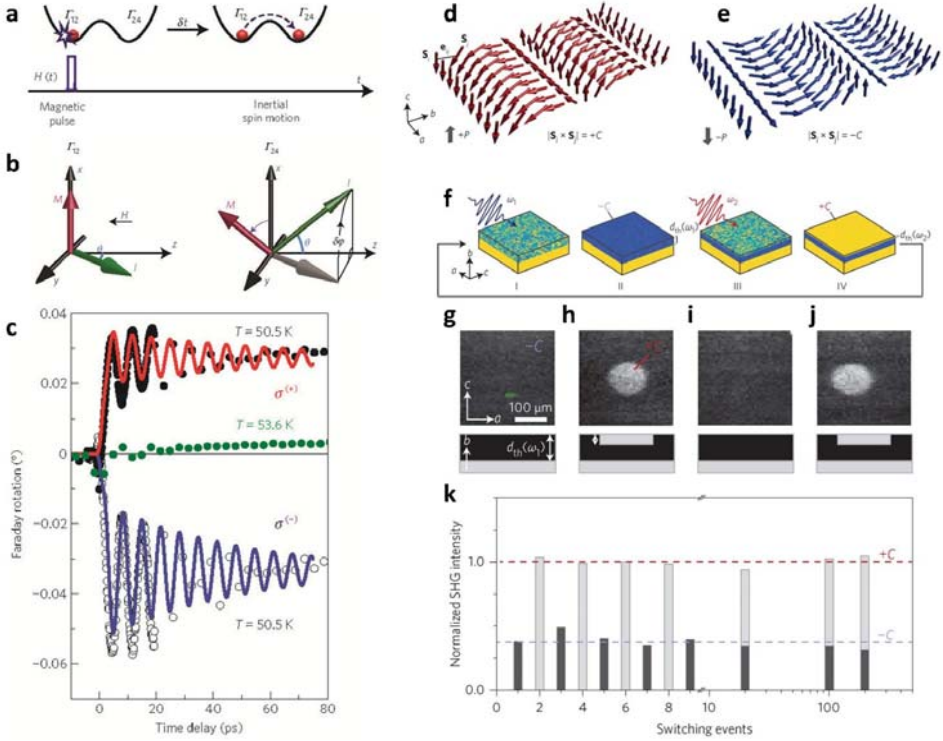

**Figure 4 Optical switching of the antiferromagneticstate in HoFeO$_3$ (a-c) and TbMnO$_3$ (d-k).**
**a**, Schematic of the inertial switching between two metastable states. During the action of the driving force the coordinate of the particle is hardly changed, but the particle acquires enough momentum to overcome the barrier afterwards. **b**, HoFeO$_3$ has two metastable phases $\Gamma_{12}$ and $\Gamma_{24}$ present between 38 and 52 K. **H** represents an effective magnetic-field pulse with a duration of 100 fs that initiates an inertial motion of spins from the $\Gamma_{12}$ to the $\Gamma_{24}$ magnetic phase. In the $\Gamma_{12}$ phase the antiferromagnetic vector **L** is in the $yz$ plane and the $z$ component of magnetization **M** is zero. A transition towards the $\Gamma_{24}$ phase occurs through a rotation of **L** over an angle $\delta\varphi$ towards the new equilibrium in the $zx$ plane with a non-zero $M_z$ component. **c**, The effective magnetic fields along the $z$-axis were generated by right-handed ($\sigma^+$) and left-handed ($\sigma^-$) circularly polarized laser pulses through the inverse Faraday effect. The trace recorded at $T = 53.6$ K for a $\sigma^-$ pump is shown by green points and demonstrates that outside the range 38-52 K no inertia-driven spin switching is seen.

**d, e**, Antiferromagnetic order in TbMnO$_3$ with spin cycloids of opposite helicity ±C and associated spin-induced electric polarization $\pm P \propto \pm C$. **f**, Schematics of reversible two-colour switching. (I) Ilumination with $\omega_1$ light heats a region of thickness $d_{th}(\omega_1)$ above the ordering temperature $T_0$. The underlying bulk remains in the +C state (yellow). (II) After re-cooling, the dipolar stray field exerted by the uniformly polarized environment, which was not photoexcited above $T_{AFM}$, forced the cooling region with a thickness $d_{th}(\omega_1)$ into the opposite polarization state −C state (blue). (III) Illumination with $\omega_2$ light heats a region $d_{th}(\omega_2) < d_{th}(\omega_1)$ above $T_0$. The intermediate layer remains in the −C state (blue). (IV) Because the intermediate −C layer (blue) screens the stray field of the bulk state (yellow), the $d_{th}(\omega_2)$ top layer has reversed to the +C state (yellow). **g-k**, Experimental realization of reversible optical switching. **g**, After electric-field cooling to an antiferromagnetic +C state, illumination at $\omega_1$ reverses a top layer to −C. **h**, Local illumination at $\omega_2$ reverses the state to +C. **i**, Illumination at $\omega_1$ recreates the state in **g**. **j**, Repeat of the step in **h**. **k**, Sequential optically induced reversal of the antiferromagnetic state between +C and −C. Columns show the normalized SHG intensity in the area illuminated with light at $\omega_1$ (dark) or $\omega_2$ (bright). Figure reproduced with permission from: **a-c**, Ref. 33; **d-k**, Ref. 96.

A rather different approach was used in Ref. 96 to achieve reversible optical switching in TbMnO$_3$. In this material, a helimagnetic spin cycloid with zero net magnetization arises from competing magnetic interactions in the Mn$^{3+}$ sublattice below $T_{AFM}$ = 27 K and it induces a ferroelectric polarization **P**, which can be used as a handle to control the antiferromagnetic order [see Figs. 4(d) and (e)][96]. Finally, a distinct approach was used in metallic CuMnAs where a reversible switching of antiferromagnetic domains via a spin-orbit torque was induced using a picosecond long electric field transient of a THz-band radiation whose polarization direction determined the domain orientation[141] (For more details see the article on spin-transport, spin-torque and memory of this focused issue).

IV. PRECESSION OF MAGNETIC MOMENTS

In this chapter, we describe mechanisms that can be used for inducing the precession of spins in AFs. The generation and detection of spin precession by electromagnetic radiation is very important also for antiferromagnetic material development and optimization because these methods can serve as a tool for measuring the antiferromagnetic resonance[44]. In fact, spin precession frequencies in AFs are so high that it is extremely difficult to measure them by common magnetic resonance techniques. For example, for spin waves at the edge of the Brillouin zone of KNiF$_3$, a 20 THz frequency was reported[35]. Moreover, optical techniques enable to obtain a simultaneous high time- and spatial-resolutions, limited only by the duration and wavelength of the used pulses, respectively. Finally, as we demonstrate below, spin precession can be also used as a means for information storage.

There are several mechanisms that can be used to induce optically precession of magnetic moments in AFs. Conceptually, the most straightforward mechanism is the Zeeman torque due to direct coupling between the magnetic field component of an electro-magnetic wave and the spins. Intense pulses of THz radiation with a duration shorter than 1 ps and electric- and magnetic-field amplitudes exceeding 1 MVcm$^{-1}$ and 0.33 T, respectively, have become available recently[142,110]. In Figs. 5(a) and (b) we show experimental data demonstrating this effect in NiO (Ref. 28) and in Figs. 5(c) and (d) in YFeO$_3$ (Ref. 120). Similar results were also reported by other groups in YFeO$_3$ (Ref. 121) and DyFeO$_3$ (Ref. 47). Compared to the more common indirect coupling of the electric field component of the light with spins via spin-orbit interaction, the magnetic component of a THz pulse exclusively addresses the electron spins and does not deposit excess heat in other degrees of freedom. For example, it was estimated in Ref. 28 that the energy deposited by the THz pump pulse increases the temperature of the excited NiO volume by less than 10 mK. THz magnetic fields are, therefore, ideally suited for ultrafast coherent control experiments where sequences of pulses are used to switch the precession modes on and off [see Fig. 5(d) and Fig. 3 in Ref. 28]. In addition to Zeeman coupling, intense terahertz electric fields can couple to magnetic excitations nonlinearly through electric-dipole transitions that modify the anisotropy field [see Figs. 5(e)-(i)][48].

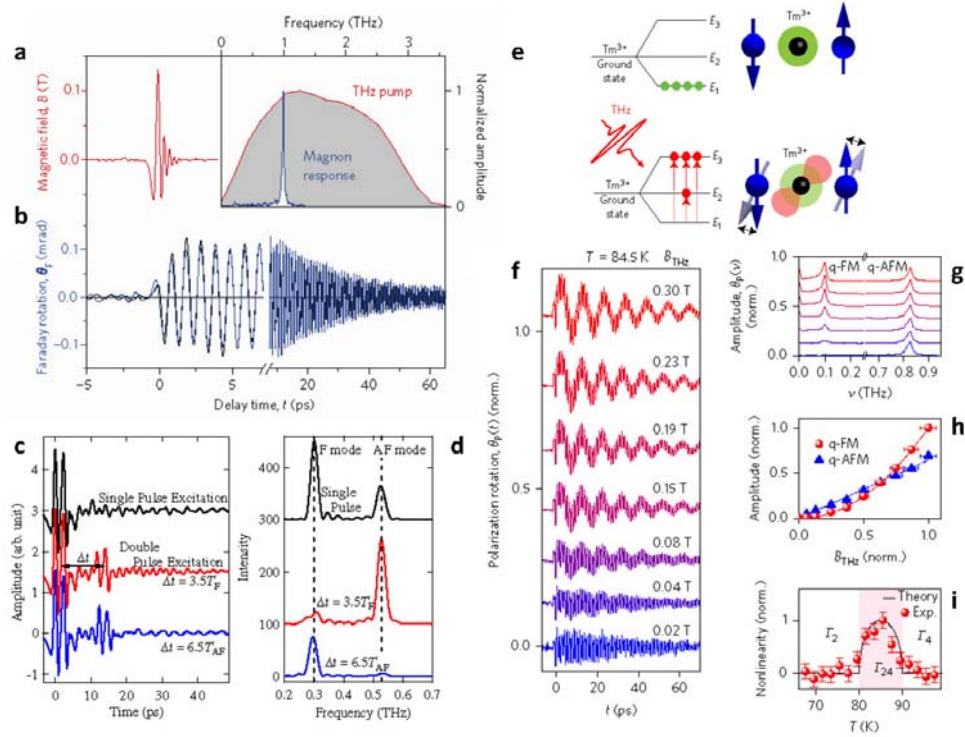

**Figure 5 Magnetization precession induced by THz pulses in NiO (a-b), YFeO$_3$ (c-d), and TmFeO$_3$ (e-i).**
**a**, Magnetic field of the incident terahertz pulse as a function of time. **b**, Ultrafast Faraday rotation in the NiO sample (blue curve: experiment; black line: simulation); harmonic oscillations with a period of 1 ps are due to an antiferromagnetic spin precession. Inset: amplitude spectra of the Faraday transient (**b**) and the driving terahertz field (**a**).

**c**, Temporal waveforms of THz pulse transmitted through an *a*-cut plate of YFeO$_3$ for single-pulse and double-pulse excitations. Two pulses are separated in time by time delay $\Delta t$, $T_F$ and $T_{AF}$ are the oscillation period of the ferromagnetic and antiferromagnetic mode, respectively. **d**, Spectra obtained by Fourier transformation from 18 to 48 ps in **c** which shows that the precession modes can be canceled independently.

**e**, The crystal field splits the ground state $^3H_6$ of the rare-earth Tm$^{3+}$ ions into several energy levels with an energy spacing of 1-10 meV (schematic level scheme). The corresponding orbital wavefunctions set the magnetic anisotropy for the Fe$^{3+}$ spins in thermal equilibrium (upper panel). Ultrafast transitions between these energy levels induced by resonant terahertz pulses abruptly modifies the magnetic anisotropy and exerts torque on the spins, which triggers coherent spin dynamics (lower panel). **f**, Normalized magnon traces measured for various terahertz excitation fields $B_{THz}$. While quasimonochromatic oscillations are found for the lowest terahertz field, a low-frequency oscillation is superimposed onto the dynamics for higher pump fields. **g**, Amplitude spectra of the time-domain data shown in **f** allow the identification of quasi-ferromagnetic (q-FM) and quasi-antiferromagnetic (q-AFM) modes at 100 and 830 GHz, respectively. **h**, Scaling of the amplitudes from **g**. The q-AFM mode scales linearly with the terahertz field strength, whereas the q-FM mode shows a nonlinear dependence. **i**, Deviation of the experimental field-scaling of the q-FM mode with terahertz field strength from a linear behaviour (points) and computed terahertz-induced anisotropy torque exerted on the spins (curve). Figure reproduced with permission from: **a-b**, Ref. 28; **c-d**, Ref. 120; **e-i**, Ref. 48.

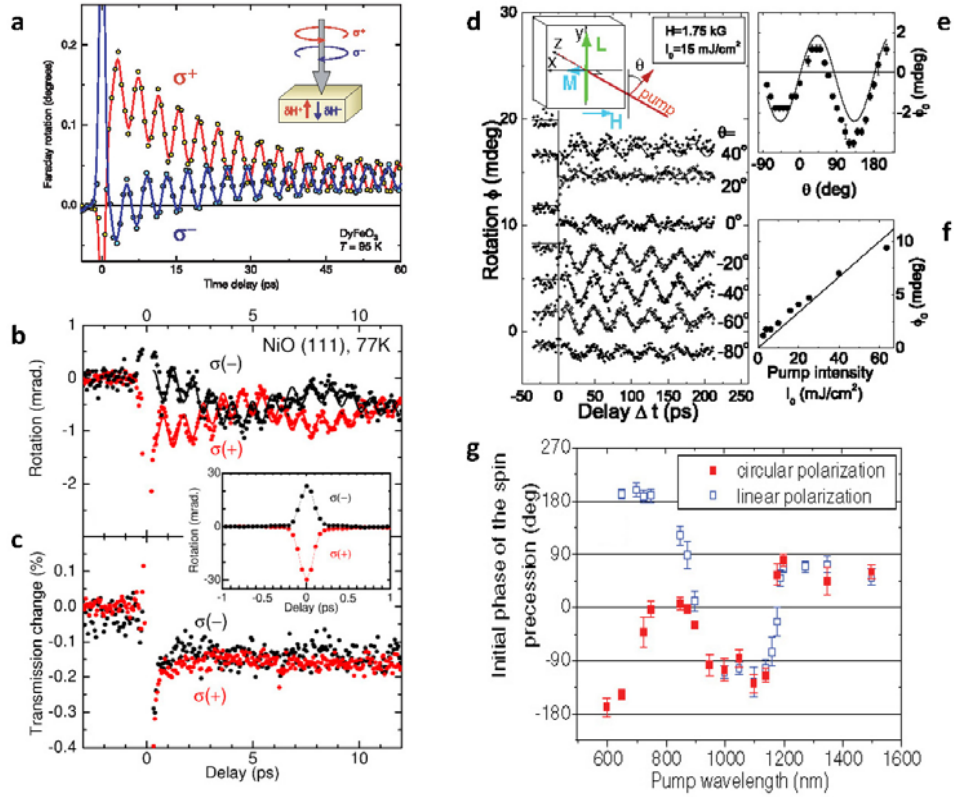

**Figure 6 Magnetization precession induced by inverse magneto-optical effects in DyFeO$_3$ (a, g), NiO (b-c), and FeBO$_3$ (d-f).**

**a**, Oscillations of the polarization rotation measured in canted AF DyFeO$_3$ due to precession of Fe spins. The circularly polarized pumps of opposite helicities excite oscillations of opposite phase. Vectors δ**H**$^+$ and δ**H**$^-$ represent the effective magnetic fields induced by right-handed ($\sigma^+$) and left-handed ($\sigma^-$) circularly polarized pumps, respectively. **b** and **c**, Time-resolved polarization rotation (**b**) and transmission change (**c**) measured in compensated AF NiO (111) for $\sigma^+$ and $\sigma^-$ pump helicities. **d**, Time-resolved polarization rotation measured in FeBO$_3$ for different linear polarizations of the pump (see inset). **e, f**, The oscillation amplitude *vs* orientation and intensity of linearly polarized pump, respectively. **g**, Initial phase of the oscillation as a function of the pump wavelength in DyFeO$_3$. For wavelengths shorter than ≈ 800 nm, when the phase is a multiple of π, the inverse Faraday effect dominates while for longer wavelengths, when the phase is an odd multiple of π/2, the inverse Cotton-Mouton effect dominates[45]. Figure reproduced with permission from: **a**, Ref. 43; **b-c**, Ref. 27; **d-f**, Ref. 45; **g**, Ref. 46.

Magnetic anisotropy can also be modified by an optically induced temperature increase [see Figs. 3(j) and (k)], which was responsible for the precession reported in TmFeO$_3$ (Refs. 42 and 44) and DyFeO$_3$ (Ref. 131). Alternatively, inverse magneto-optical effects can induce the precession in transparent magnetic materials with a strong MO response. The inverse Faraday effect was theoretically predicted in Ref. 143 and subsequently experimentally observed in Ref. 144. In Figs. 6(a) and (b) we show the experimental observation of this effect in canted AF DyFeO$_3$ (Ref. 43) and compensated AF NiO (Ref. 27). The fingerprint of this mechanism is a change of the precession phase with the helicity of circularly polarized pump pulses. Microscopically, this effect can be described as either optical Stark splitting of spin sublevels or as Raman scattering on magnons, depending on whether the optical field is applied parallel or perpendicular to the magnetic quantization axis[32,145]. Phenomenologically, this effect is described in terms of the action of an effective magnetic field pulse, which is present only for the duration of the optical field [see Inset in Fig. 6(a)], with an amplitude as large as 0.3 T (Ref. 43). Later, it was realized that also linearly polarized laser pulses, even though they do not carry any angular momentum, can generate effective magnetic fields through an ultrafast inverse Cotton-Mouton effect [see Figs. 6(d)-(f)][45]. Its fingerprint is the harmonic dependence on the linear polarization orientation [cf. Eq. (2) and Fig. 6(e), see also Ref. 52]. The inverse Faraday and Cotton-Mouton effects induce oscillation with a distinct phase that can be used for their experimental separation if they coexist in a studied material such as $DyFeO_3$ (Ref. 46) [see Fig. 6(g)] and NiO (Ref. 52). Another mechanism, which can induce precession of spins in a broad class of iron oxides with canted spin configuration (for example in $FeBO_3$, $TmFeO_3$, $YFeO_3$, and $ErFeO_3$), is the inverse magneto-refraction. The underlying principle is the modification of the exchange interaction by the electric field of light, which for laser pulses with a fluence of about 1 mJcm$^{-2}$ acts as a pulsed effective magnetic field with a magnitude of 0.01 T[116]. Similarly, optical modification of *d-f* exchange interaction between conduction band electrons and lattice spins was shown to trigger spin waves in EuTe[60].

Potential application of magnetic oscillations is in information storage. Traditionally, optical magnetization control has been limited to a binary process, where light in either of two polarization states writes or reads a magnetic bit carrying either a positive or negative magnetization[146]. However, it is possible to achieve a full vectorial control of magnetization by light that can be used for storing multiple pieces of information in a single storage element[4,113]. The implementation requires to independently address the phase and amplitude of multiple degenerate magnetization modes. This was done in antiferromagnetic NiO (111), where a micro-multidomain structure with two selectively excited magnetic modes enabled full control of two-dimensional magnetic oscillations with a pair of time-delayed polarization-twisted femtosecond laser pulses [see Figs. 7(a)-(e)][113] Later, this approach was extended to a full three-dimensional control of magnetic oscillations in the three-sublattice antiferromagnet $YMnO_3$ [see Fig. 7(f)][4]. The idea relies on a one-to-one mapping of the three Stokes parameters, which parameterize the light polarization on the Poincaré sphere, onto three magnetic oscillation modes [see Figs. 7(g)-(k)][4]. It was also demonstrated that the magnetic oscillation state can be transferred back into the polarization state of an optical probe pulse, thus completing an arbitrary optomagnonic write–read cycle[4].

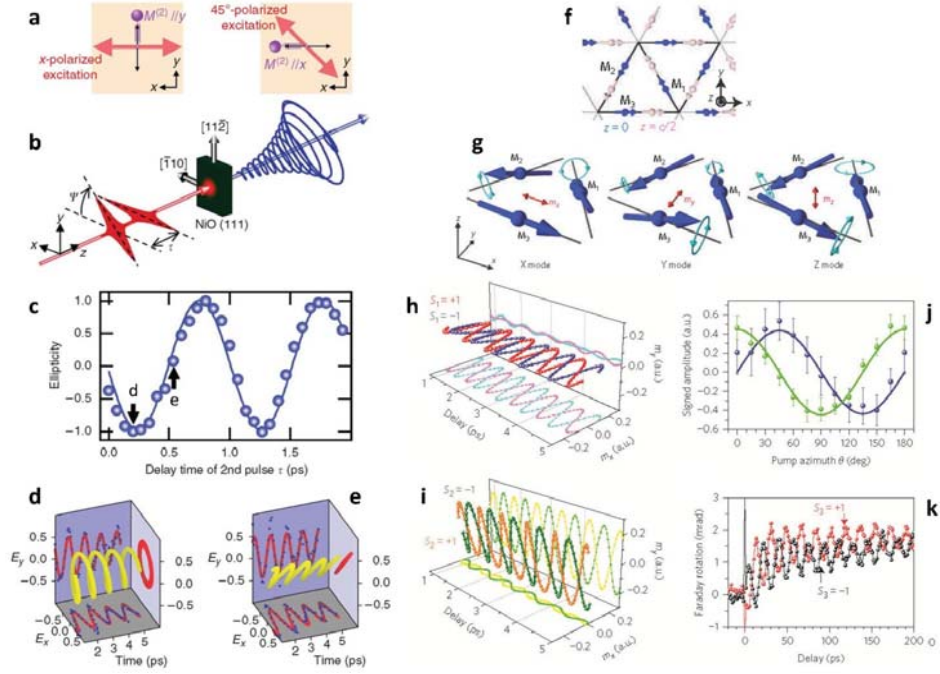

**Figure 7 Vectorial control of magnetization by light in NiO (a-e) and YMnO$_3$ (f-k).**
**a**, Two orthogonal magnetic oscillation modes in NiO crystal that can be selectively excited by tuning the polarization azimuth of the excitation light pulse. **b**, Schematic illustrations of the vectorial control of the magnetization vector with polarization-twisted double-pulse excitation. Both pump pulses are linearly polarized and have a relative time delay $\tau$. The first pulse is polarized along the *x*-axis, whereas the second pulse has an azimuthal angle of $\psi$. By properly tuning $\tau$ and $\psi$ the motion of the magnetization vector can follow an arbitrarily designed direction and amplitude of polarization. **c**, Ellipticity of the measured THz radiation at $\Omega_{\mathrm{mag}}$ as a function of dealy $\tau$ between the two linearly polarized excitation pulses for $\psi = 45°$; any ellipticity between the purely right and left circular polarizations is obtained by tuning $\tau$. The solid curve is a fit of the data by a sine curve. The three-dimensional trajectories of the electric field vectors with fixed values of $\tau$ lead to purely circularly and linearly polarized radiation as shown in **d** and **e** respectively. The dots correspond to the projections of the experimental data onto the *x*- and *y*-axes. The solid curves correspond to the fit of exponentially decaying sine functions; the phase differences between the *x*- and *y*-components are determined by $\Omega_{\mathrm{mag}}\tau/\pi$.

**f**, Antiferromagnetic three-sublattice ordering of the magnetic Mn$^{3+}$ moments in hexagonal YMnO$_3$. **g**, Magnetic oscillation eigenmodes X, Y and Z. **h**, Magnetic 1.3 THz oscillation components excited by a pump pulse with a polarization described by a Stokes parameter $S_1= \pm 1$ (i.e., pump is linearly polarized along *x*- or *y*-axes, respectively). **i**, Same as **h** for $S_2=\pm 1$ (i.e., pump is linearly polarized along diagonals (*x+y*) and (*x-y*), respectively). **j**, Signal detected in the X-probe (green) and Y-probe (blue) configurations, as defined in Ref. 4, as a function of the angle $\theta$ parameterizing the orientation of pump pulse linear polarization. Lines are proportional to cos $2\theta$ (green) and sin $2\theta$ (blue). **k**, Time-dependent Faraday rotation exerted on the probe pulse after excitation by a pump pulse with a polarization described by the Stokes parameter $S_3=\pm 1$ (i.e., pump is circularly $\sigma_+$ or $\sigma_-$ polarized, respectively). Figure reproduced with permission from: **a-e**, Ref. 113; **f-k**, Ref. 4.

OUTLOOK

Methods based on the interaction of electromagnetic radiation with magnetically ordered materials have been essential for a research of AFs in the past and have a large potential to contribute to this rapidly evolving scientific field also in the future. For instance, due to high frequencies of spin excitations, AFs are considered as appealing materials for future THz spintronics. However, many concepts are yet to be demonstrated including, e.g., the ultrafast spin pumping from AFs. Experimental investigation of the spatial and temporal evolution of spin pumping from AFs is therefore one of the next challenges in the field of antiferromagnetic opto-spintronics. A similar situation is found in the emerging field of THz opto-magnonics. It has been demonstrated that light can effectively generate and control spin waves at THz frequencies, but propagation of these spin excitations in space has not been studied yet and represents another challenge in ultrafast magnetism. AFs with non-collinear spin alignment offer a plethora of opportunities for further development of ultrafast opto-magnetism and opto-spintronics. The methods described in this review can be employed for investigation of yet unexplored dynamics and mechanisms of the control of antiferromagnetic skyrmions. For multiferroics, which are often AFs, dynamical phenomena are still a highly underrated topic. With magnetoelectric switching as one of the declared goals of the field, hardly any attention is paid to the temporal evolution of these reorientations. One important aspect is the speed of the reversal of the antiferromagnetic and ferroelectric order parameter, another one is the evolution of the corresponding domain patterns.

Acknowledgements

PN acknowledges support from the Grant Agency of the Czech Republic under Grant No. 14-37427G and by the Ministry of Education of the Czech Republic Grant No. LM2011026. AVK acknowledges the Netherlands Foundation of Scientific Research (NWO). TK thanks the European Research Council for support through Grant No. 681917 (TERAMAG). MF acknowledges support from the SNSF project 200021/147080 and by FAST, a division of the SNSF NCCR MUST.